\documentclass[journal,comsoc]{IEEEtran}
\usepackage{graphicx}
\usepackage{array}
\usepackage{color}
\usepackage{epsf}
\usepackage{times}
\usepackage{epsfig}
\usepackage{graphicx}
\usepackage{epstopdf}
\usepackage{hyperref}
\usepackage{cite}
\usepackage{amsmath}
\usepackage{amssymb}
\usepackage{amsxtra}
\usepackage{amsthm}
\usepackage{bbm}
\usepackage{algorithmic}
\usepackage{algorithm}
\usepackage{subfigure}
\usepackage{authblk}
\usepackage{url}
\usepackage{comment}
\usepackage{siunitx}
\usepackage{lipsum}
\usepackage{kantlipsum}
\usepackage{dblfloatfix}
\usepackage{adjustbox}
\usepackage{tabularx}
\usepackage{multirow}
\usepackage{float}
\usepackage{framed}
\usepackage{caption}
\begin{document}
\title{Full-Duplex Magnetic Induction Communication: Opportunities and Challenges}
\author{Muhammad Muzzammil, Saif Al-Kuwari,~\IEEEmembership{Senior Member,~IEEE,} Niaz Ahmed,~\IEEEmembership{Member,~IEEE,} Marwa Qaraqe~\IEEEmembership{Senior Member,~IEEE}

\thanks{ This work is funded by the G5828 ``SeaSec: DroNets for Maritime Border and Port Security" project under the NATO's Science for Peace Programme.  }
\thanks{M. Muzzammil is with the Division of Information and Computing Technology, College of Science and Engineering, Hamad Bin Khalifa University, Qatar Foundation, Doha, Qatar and the College of Underwater Acoustic Engineering, Harbin Engineering University, China, Email: muzzammilm@hrbeu.edu.cn.\newline
S. Al-Kuwari and M. Qaraqe are with the Division of Information and Computing Technology, College of Science and Engineering, Hamad Bin Khalifa University, Qatar Foundation, Doha, Qatar, Email: \{smalkuwari, mqaraqe\}@hbku.edu.qa.\newline
N. Ahmed is with the Department of Electrical Engineering, FAST National University of Computer and Emerging Sciences, Islamabad Campus, Email: niaz.ahmed@nu.edu.pk}
}

\maketitle

\begin{abstract}
The demand for high data rates is rapidly increasing as the interest in Magnetic Induction (MI) communication-based underwater applications grow. However, the data rate in MI is limited by the use of low operational frequency in generating a quasi-static magnetic field. In this paper, we propose the use of full-duplex (FD) MI communication to efficiently utilize the available bandwidth and instantly double the data rate. We propose a two-dimensional transceiver architecture to achieve full-duplex communication by exploiting the directional nature of magnetic fields. We further evaluate the proposed end-to-end FD MI communication against self-interference (SI), its impact on communication distance, and robustness in view of orientation sensitivity. Finally, we conclude by discussing typical challenges in the realization of FD MI communication and highlight a few potential future research directions.
\end{abstract}

\IEEEpeerreviewmaketitle

\section{Introduction}
\label{sec:intro}
Recent advances in UWSNs have led to the development of underwater vehicles, such as autonomous underwater vehicles (AUVs) and unmanned underwater vehicles (UUVs). This has naturally unfolded many underwater applications starting from  the exploration of natural resources and monitoring of the underwater environment, to real-time military surveillance~\cite{li2019survey}. Consequently, demand for high data rates has increased, especially for real-time applications involving the transmission of high-resolution images and videos~\cite{kida2018experimental, wang2018high}. Hence, this challenges the research community to explore technologies and solutions that can not only provide high data rates but also efficiently utilize the available bandwidth in an underwater environment. 
Among the three popular modes of underwater wireless communications: acoustic, optical and magnetic induction (MI) communication, serious efforts are already made in acoustic and optic-based sensor networks to increase data rate and exploit the spectral efficiency~\cite{songzuo2021full, li2020real}. Similar efforts are thus required in MI wireless sensor networks as well to explore and fully exploit their potential. However, in MI communication, low operational frequency is generally used to generate a quasi-static magnetic field, which puts an upper limit on data rates. Therefore, in this paper, we propose the use of FD MI communication, which can instantly double the data rate.

Full-duplex communication enables two-way exchange of information simultaenously at a given frequency~\cite{sabharwal2014band}. This means that FD communication can maximize spectral efficiency and double the data rate. Apart from enhancing the spectral efficiency at the physical layer, full-duplex can also improve the efficiency at lower layers, i.e. medium access control (MAC), by introducing FD frame levels, potentially enabling sensor nodes to transmit and receive incoming and outgoing frames simultaneously. Sensor nodes with such capabilities can be helpful in both detecting collisions and receiving immediate feedback from other sensor nodes~\cite{sabharwal2014band}.
However,  implementation of any FD communication system is challenging due to self-interference (SI), synchronization, attenuation, noise, multipath, and Doppler effect. In case of underwater channel, these challenges become even more severe, but due to simple physical layer modeling of MI communication, factors like synchronization, multipath, and Doppler are already suppressed. 

\begin{figure}[]
    \centering
    \includegraphics[width=0.49\textwidth]{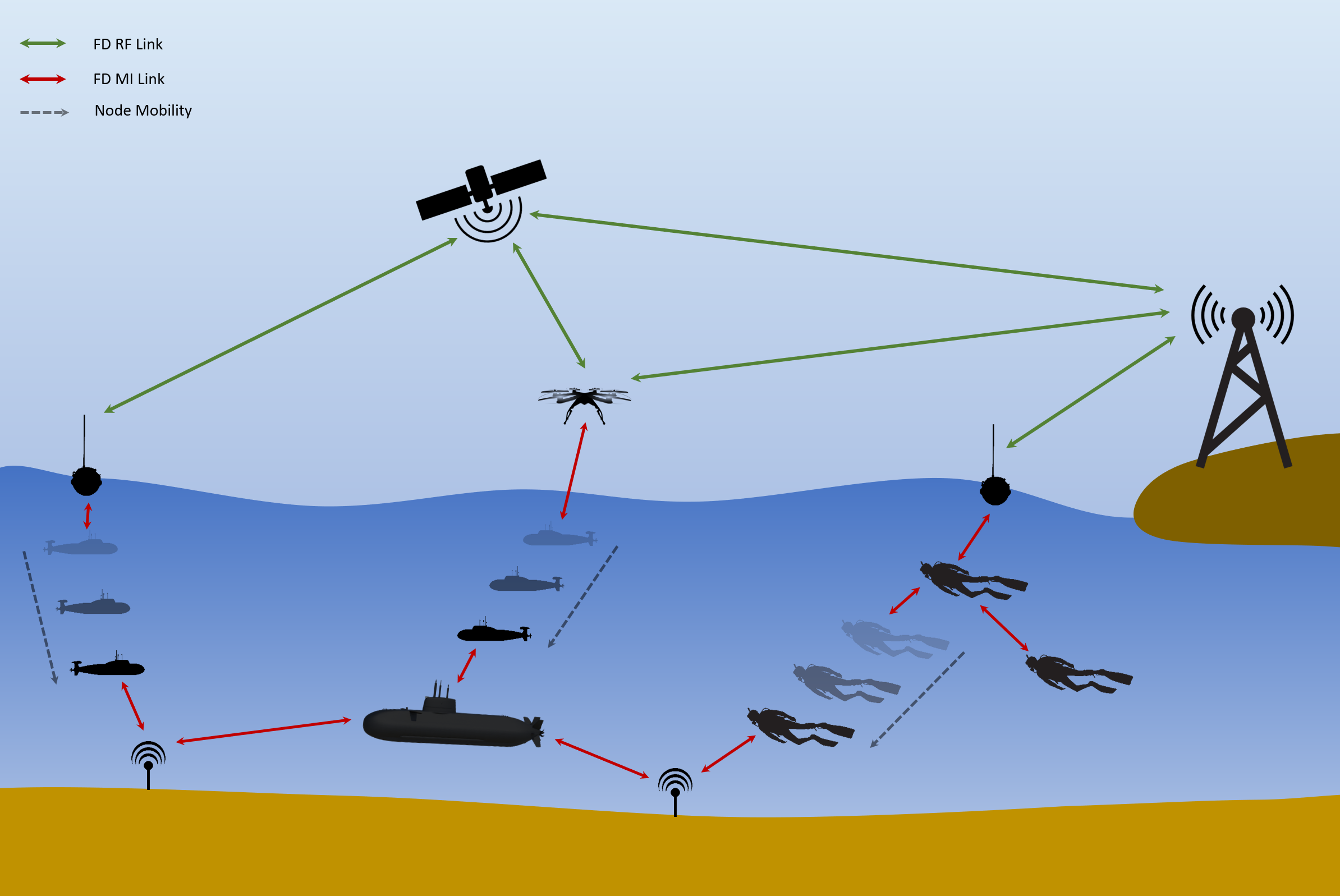} 
    \caption{General architecture of FD MI-UWSNs.}
    \label{fig:illustration}
    \end{figure}

\begin{figure*}[]
 %\begin{framed}
% \vspace{-0.50cm}
 \centering
    \subfigure[\label{subfig:FD_architecture}]{\includegraphics[width=1\textwidth]{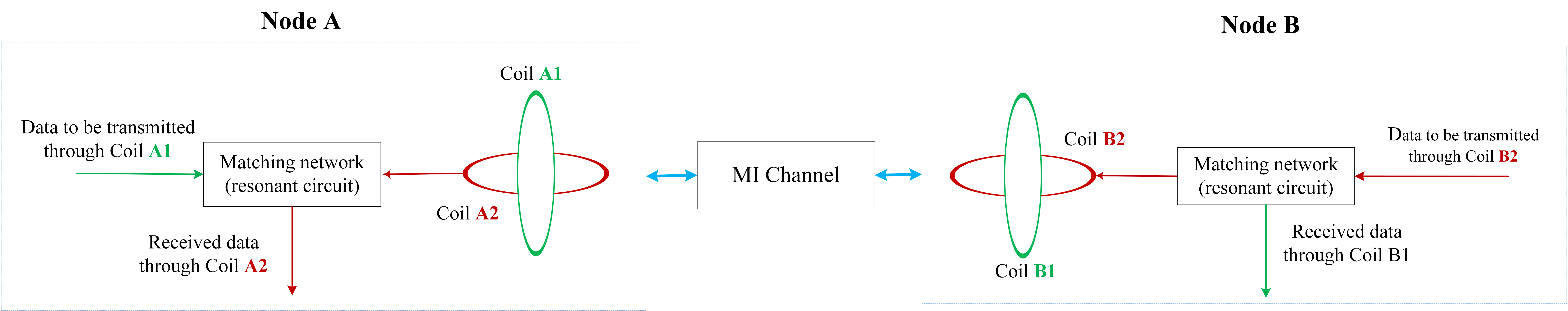}} 
  \subfigure[\label{subfig:FD_circuit.}]{\includegraphics[width=1\textwidth]{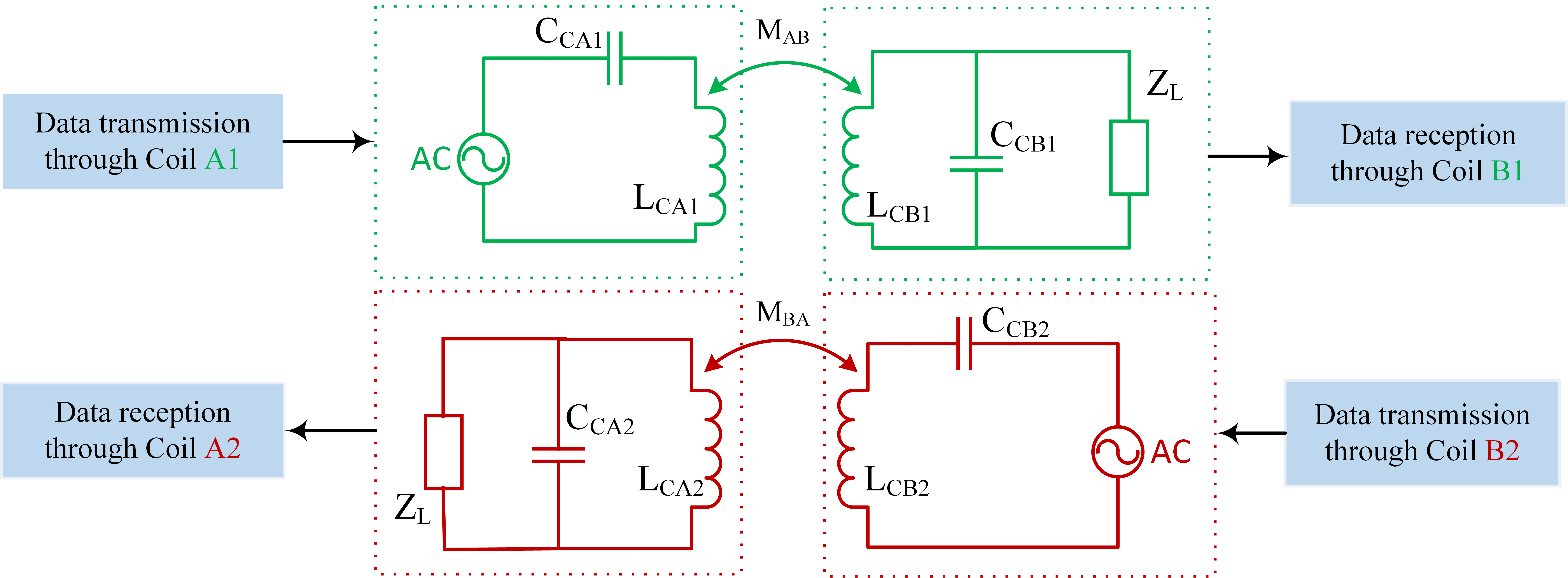}}
   %\vspace{-0.2cm}
 \caption{Full-duplex MI communication: (\textbf{a}) General architecture, and (\textbf{b}) Equivalent circuit model.}
 \label{fig:FD_MI}
 %\vspace{-0.2cm}
 %\end{framed}
 \end{figure*}

In this paper, we discuss an end-to-end FD MI communication system and highlight the major challenges in exploiting these opportunities, such as SI, communication distance, and orientation sensitivity. To the best of our knowledge, only a few existing studies considered FD MI communication in an underwater environment. In~\cite{guo2018full}, the authors proposed a meta-material enhanced MI network with a FD relay node to extend the communication distance. On the other hand, in this paper, we present a simple and novel transceiver design that is capable of FD MI communication. Our proposed design exploits the directional nature of the magnetic fields and is explained in detail in the following section. We also evaluate the proposed end-to-end FD MI communication architecture based on self-inference, communication distance, and orientation sensitivity.

The rest of the paper is organized as follows: section~\ref{sec:FD_MI} presents the architecture of our proposed end-to-end FD MI communication followed by the discussion on its potential and system evaluation. Section \ref{sec:FD_challenges} discusses some exciting research challenges in FD MI communication and future directions. Finally, section~\ref{sec:conclusion} concludes the paper.
 
%%%%%%%%% SECTION II : FD MI Communication %%%%%%%%%%%

\section{Full-Duplex MI Communication: Opportunities}
\label{sec:FD_MI}
In this section, we discuss the opportunity of using a pair of two-dimensional orthogonal coils as an end-to-end FD MI communication for MI-UWSNs applications as shown in Fig.~\ref{fig:illustration}. Furthermore, we evaluate the performance of the proposed coil architecture against SI, its impact on the communication distance, and robustness considering the orientation sensitivity.

\subsection{Architecture of FD MI Communication}
\label{subsec:FD_architecture}
In a conventional MI communication system, a pair of simple low-cost copper wounded coils are used as both transmitter (Tx) and receiver (Rx). At the Tx end, a time-varying signal (current/voltage) is applied to the coil to generate a time-varying magnetic field around the coil. The communication is achieved when another coil (Rx) is coupled to the generated time-varying magnetic field~\cite{muzzammil2020fundamentals}. The performance of MI communication depends on various factors such as received magnetic field strength, orientation sensitivity and communication distance. The received magnetic field strength is directly related to the magnetic moment, which is the product of the number of turns, the area of the coils, the angle between the coils, and the current supplied to the transmit coil. The magnetic moment largely influences the performance of the MI communication such that the two coils couple efficiently when they are aligned on a given axis (the angle between Tx and Rx is zero). Similarly, the two coils de-couple (can not communicate) when they are orthogonal to each other (angle between Tx and Rx is $90^{\circ}$). Furthermore, the communication distance can be increased by increasing the magnetic moment and vice versa.  

\begin{figure*}[b]
% \vspace{-0.50cm}
 \centering
    \subfigure[\label{subfig:Iind_case1}]{\includegraphics[width=0.37\textwidth]{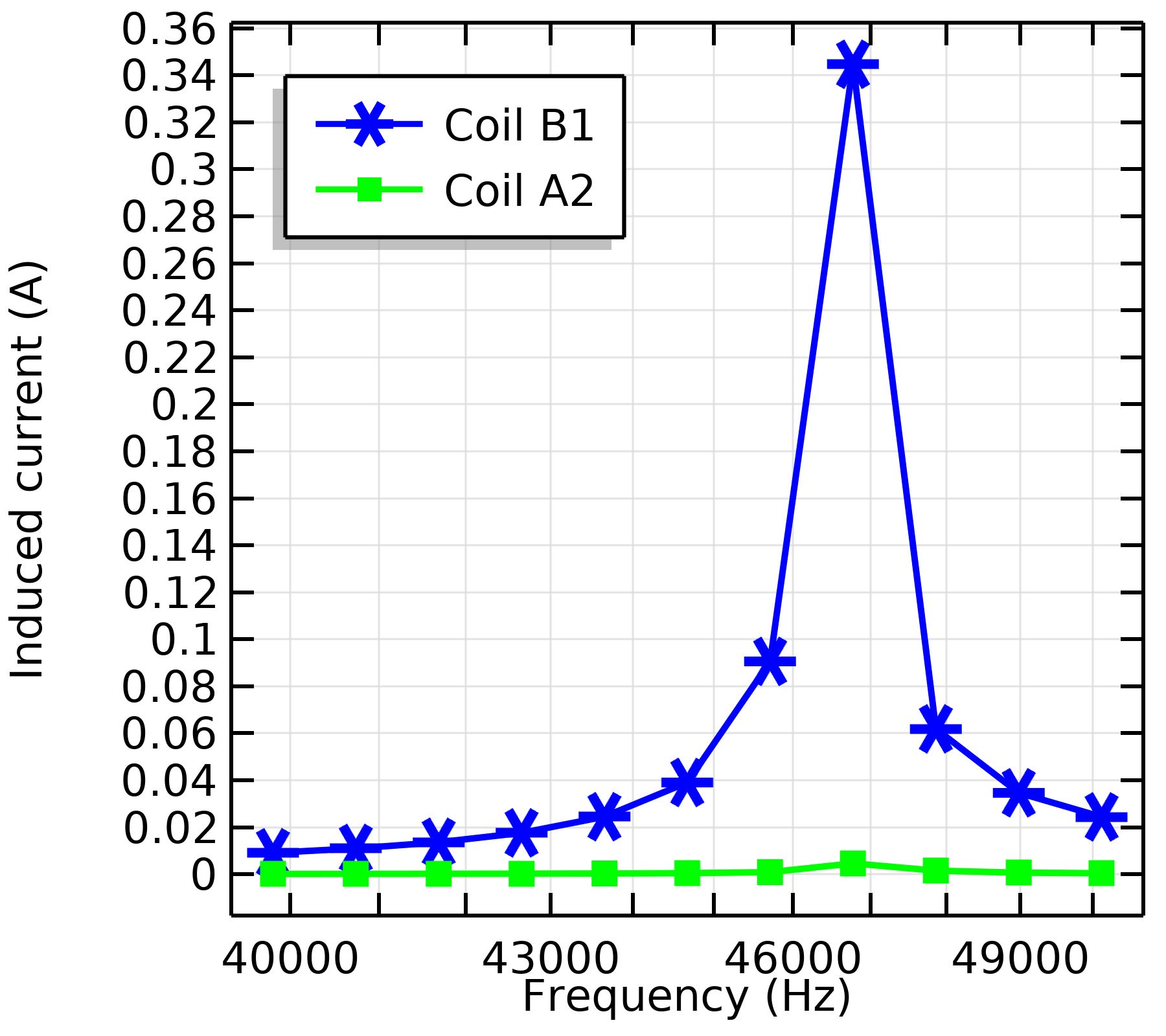}} 
  \subfigure[\label{subfig:Iind_case2}]{\includegraphics[width=0.37\textwidth]{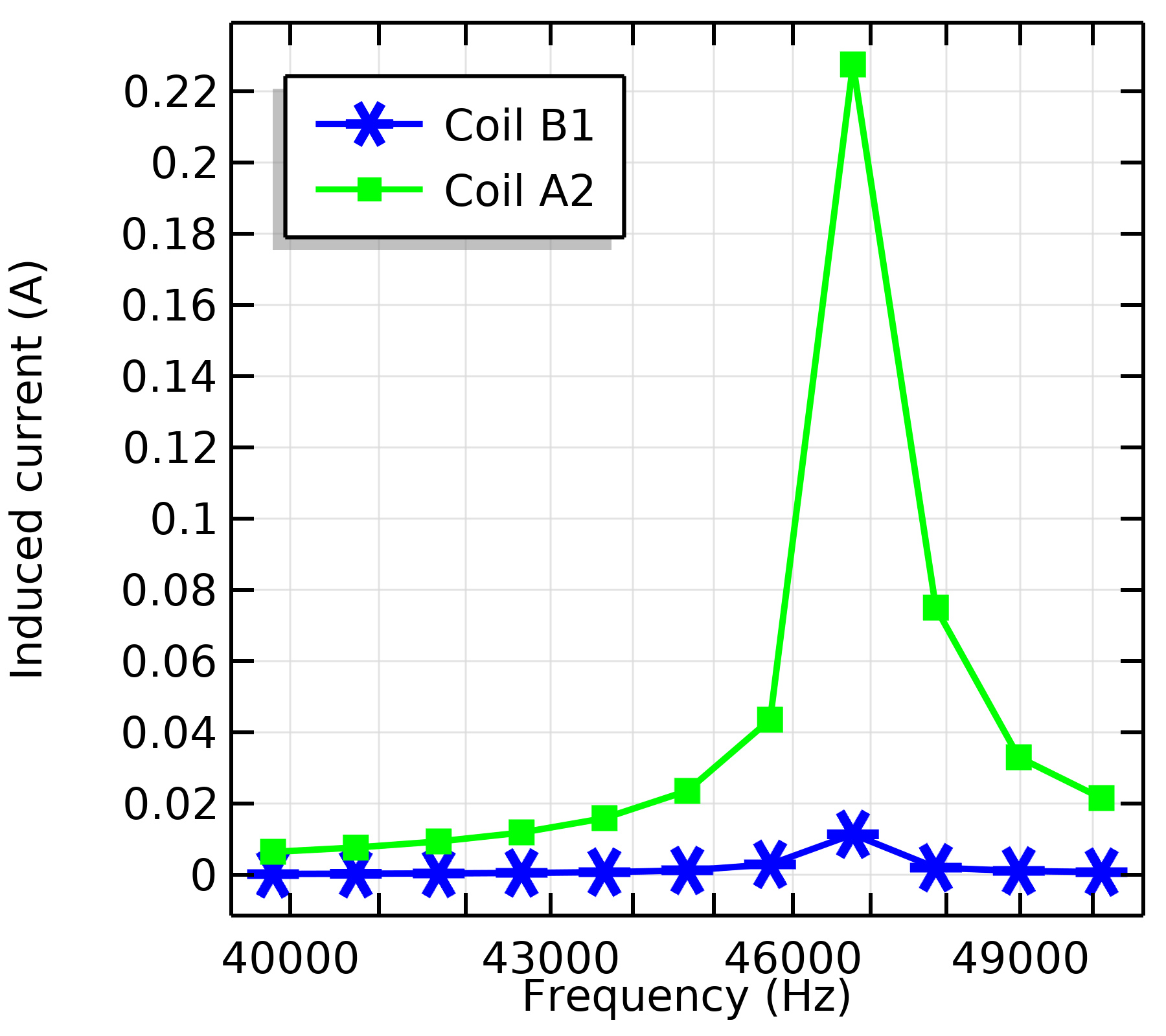}}
   %\vspace{-0.2cm}
 \caption{Induced current at the point of self-interference and intended receiver: (\textbf{a}) Only transmitter Coil A1 is excited, (\textbf{b}) Only transmitter Coil B2 is excited.}
 \label{fig:SI_case1and2}
 %\vspace{-0.2cm}
 \end{figure*}

To extend this simple Tx-Rx pair architecture, we propose a two-dimensional pair that is capable of full duplex communication. Figure~\ref{fig:FD_MI} illustrates the proposed architecture (and its equivalent circuit model) for FD MI communication. The idea is to use  Coil A1 (Tx) and Coil B1 (Rx) pair to transmit data from Node A to Node B and use Coil B2 (Tx) and Coil A2 (Rx) to transmit data from Node B to Node A, simultaneously. As shown in Fig.~\ref{fig:FD_MI}, the signal transmitted through Coil A1 is received through Coil B1 only as they are perfectly aligned (the elevation angle between Coil A1 and Coil B1 is zero). Similarly, Coil A2 and Coil B2 will not receive anything from Coil A1 as they are both orthogonal to Coil A1.  On the other hand, the signal transmitted through Coil B2 is received through Coil A2 only due to the perfect azimuth alignment of Coil B2 and Coil A2. Similarly, Coil A1 and Coil B1 will not receive anything from Coil B2 as they are both orthogonal to Coil B2.

\subsection{Performance of FD MI Communication}
\label{subsec:FD_performance}
The communication distance and orientation sensitivity of MI communications are important aspects in designing and deploying MI sensor networks, but the major challenge hindering the implementation of a full duplex communication system is self-interference. In this section, we evaluate and discuss these important factors with reference to our proposed full duplex communication system.

\subsubsection{Self-interference (SI)}
\label{subsec:SI}
In FD communication, where both the receive and transmit ends of a node are active, the node potentially receives its own transmitted signal due to direct or reflected paths. The received signal power of its own transmitted signal is generally higher than the signal of interest (SoI) from a nearby node, which causes severe self-interference (SI); therefore mitigating SI is a great challenge to any full duplex communication system. 
However, this SI is conveniently avoided using the proposed FD MI architecture exploiting the directional nature of the magnetic fields. We demonstrate this minimal effect of SI with the following three cases:

\begin{itemize}
    \item Case 1: Excite Coil A1 (to act as a transmitter), and observe the induced current/voltage at Coil A2 (point of SI) and Coil B1 (intended receiver).
    \item Case 2: Excite Coil B2 (to act as a transmitter), and observe the induced current/voltage at Coil B1 (point of SI) and Coil A2 (intended receiver).
    \item Case 3: Excite both Coil A1 and Coil B2 (the two transmitters in full duplex mode), and observe the induced current/voltage at both Coil B1 and Coil A2.
\end{itemize}

%Below we show that there is no SI in the proposed FD MI architecture or it is negligible at the system resonance frequency.

Finite element method (FEM) simulations are performed to study the three cases with the following parameters: the radius of  Coil A1 and Coil B1 is $10$cm, the radius of Coil A2 and Coil B2 is $12.1$cm, the number of turns for each coil is kept as 100, the distance between Node A and Node B is  $60$cm. Water medium (fresh water) is applied to the simulation environment with an electrical conductivity of $0.01$S/m. The coils are set to resonate at $46.77$KHz with the help of a tuning capacitor as shown in Fig.~\ref{subfig:FD_circuit.}. To resonate the complete communication system at the same frequency, the capacitors $C_{CA1}$ and $C_{CB1}$ values are kept $4.7$nF and capacitors $C_{CB2}$ and $C_{CA2}$ values are $2.45$nF. The above parameters are unchanged for each configuration unless otherwise mentioned.

\paragraph{Case 1}
In case 1, Coil A1 is excited with $1$A current while the induced current is observed at both Coil A2 and Coil B1. Induction of current in Coil A2 in this case would imply the presence of SI whereas induction of current in Coil B1 would imply reception of the SoI.

Figure \ref{subfig:Iind_case1} shows the  induced current at both Coil A2 and Coil B1 against different frequencies. Maximum induction occurs at the resonance frequency ($46.77$ KHz) of the system. Reception of the SoI can thus be clearly seen on Coil B1 whereas little to no SI is observed at Coil A2. Hence, it can be concluded that the proposed end-to-end FD MI communication system is immune to SI and therefore can be easily and practically implemented as compared to the FD mode based on acoustic or optical communication.

\paragraph{Case 2}
In case 2, Coil B2 is excited with $1$A current while the induced current is observed at both Coil A2 and Coil B1. Induction of current in Coil B1 this time  would imply the presence of SI whereas induction of current in Coil A2 would imply reception of the SoI.

Figure~\ref{subfig:Iind_case2} again shows the  induced current at both Coil A2 and Coil B1 against different frequencies. Reception of the SoI, in this case, can also be clearly seen whereas little to no SI is observed in this case too.

The two cases, therefore, confirm minimal to no SI with the proposed end-to-end FD MI communication setup.

\paragraph{Case 3}
In this case, both Coil A1 and Coil B2 are excited with $1$A current to demonstrate full-duplex communication between Node A and Node B. 

  \begin{figure}[t]
    \centering
    \includegraphics[width=0.37\textwidth]{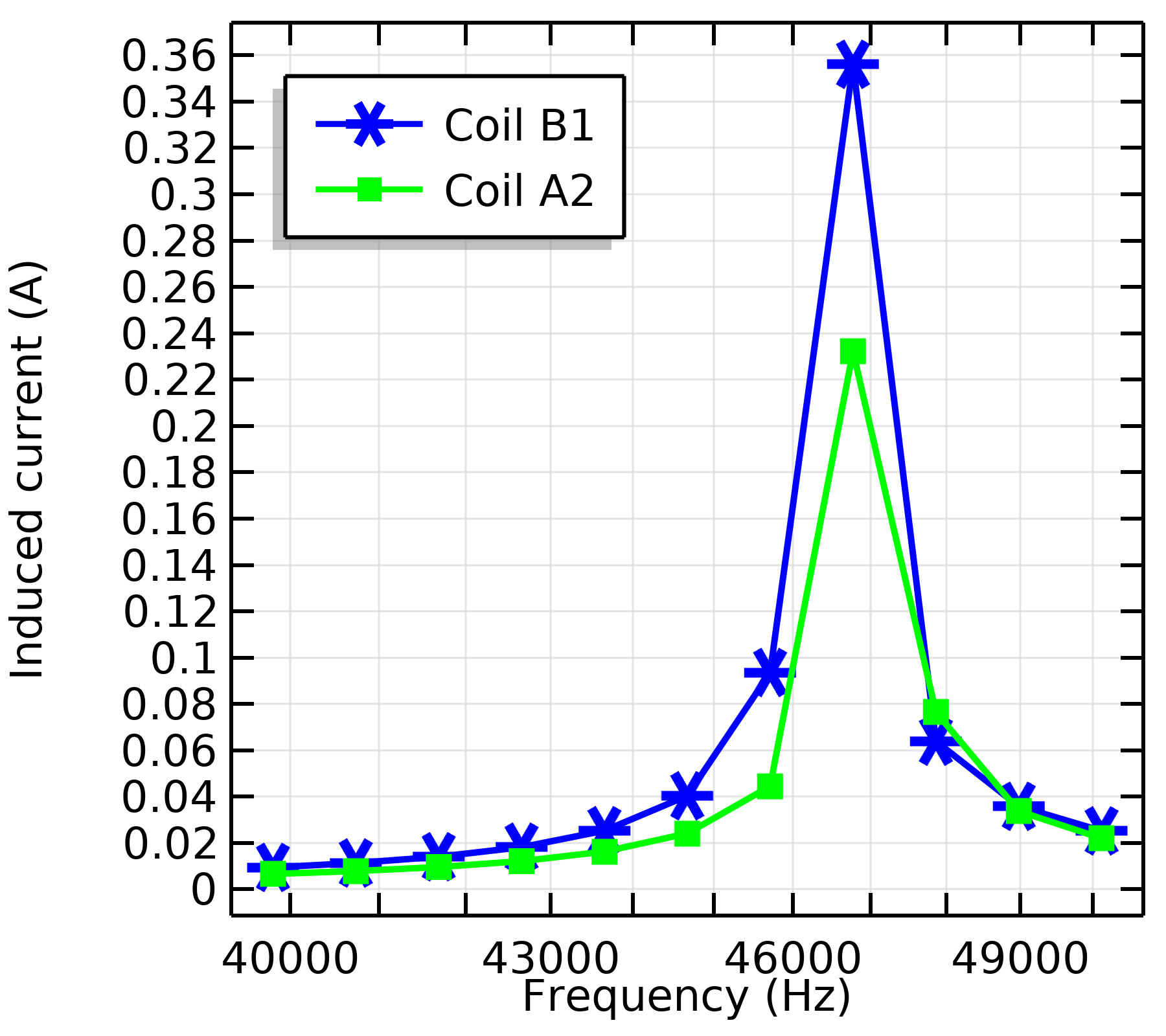} 
    \caption{Induced current in receivers (Coil B1 and Coil A2) when both transmitters (Coil A1 and Coil B2) are excited.}
    \label{fig:SI_case3}
    \end{figure}

Figure~\ref{fig:SI_case3} shows the induced current at both Coil A2 (Rx for B2) and Coil B1 (Rx for A1) against different frequencies. It can be seen that when both Coil A1 and Coil B2 are excited (transmitting a signal), Coil A2 couples to Coil B2, and Coil B1 couples to Coil A1, exactly as in case 1 and case 2. The difference in the peak values of Coil A2 and Coil B2 is due to their alignment. Since the pair A1-B1 face each other, the amount of current induced in Coil B1 is higher. Similarly, since the pair B2-A2 is placed horizontally, the induced current is relatively lower.

Considering case 1 and case 2's results, where there is minimal to no SI, the result in Fig.~\ref{fig:SI_case3} clearly demonstrates that our proposed FD MI communication system is SI-free. This unique advantage is only due to the physical nature of MI communication and can not be obtained in acoustic or optical FD communication.
    
\subsubsection{Communication Distance}
\label{subsec:range}
The proposed MI FD setup is instrumental in blocking SI, however, this setup does affect the communication distance. 

\begin{figure}[]
    \centering
    \includegraphics[width=0.37\textwidth]{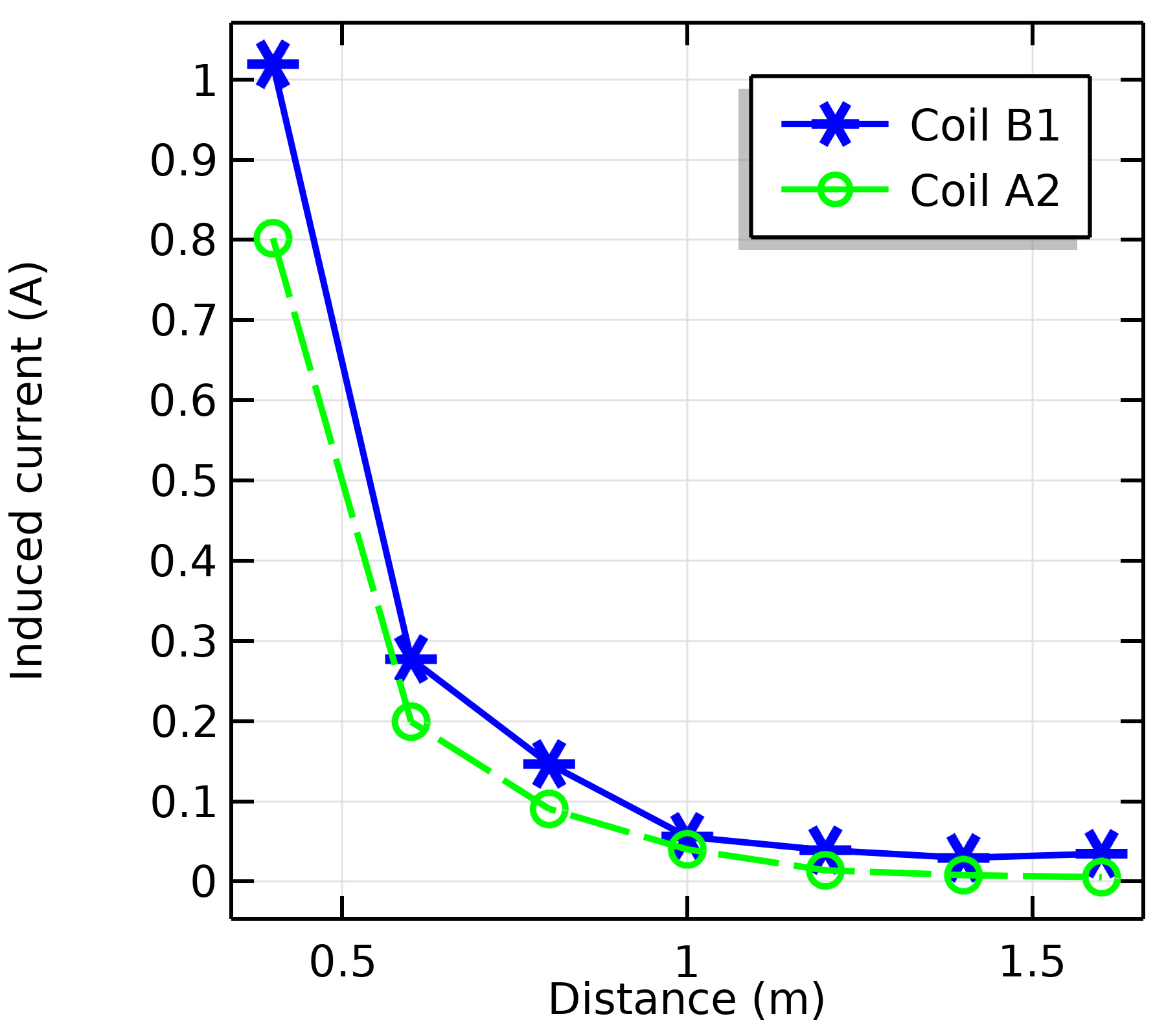}
    \caption{Induced current vs. communication distance when resonant frequency is $46.7$KHz and conductivity of water is $0.1$S/m.}
    \label{fig:mf_range}
    \end{figure}

It can be seen from Fig.~\ref{fig:SI_case3} that the induced current at the resonance frequency in the receiver Coil B1 is approx. $0.36$A, whereas the induced current at the resonance frequency in the receiver coil A2 is approximately $0.23$A. As explained earlier, the amount of current induced in the receiving end is directly related to its communication distance. 
To show the impact on the communication distance, we conducted further simulations on the proposed FD MI communication system. Fig.~\ref{fig:mf_range} shows how the induced current changes with distance. The trend in Fig.~\ref{fig:mf_range} shows that induced current is always greater in Coil B1 as compared to Coil A2. For example, at a distance of $1.6$m, the induced current in Coil B1 is non-zero while in Coil A2 it approaches zero. 

Since the induced current in the receiver Coil A2 is less than Coil B1, the communication distance will be decided based on the receiver sensitivity of Coil A2. If Coil A2's receive sensitivity is kept the same as Coil B1's receive sensitivity, it is possible that at a given communication distance, Coil B1 can still listen to Coil A1, but Coil A2 cannot listen to Coil B2. It is, therefore, strongly recommended that for the proposed setup, the receiver sensitivity of Coil A2 is larger than the receiver sensitivity of Coil B2.

\begin{figure}[]
    \centering
    \includegraphics[width=0.37\textwidth]{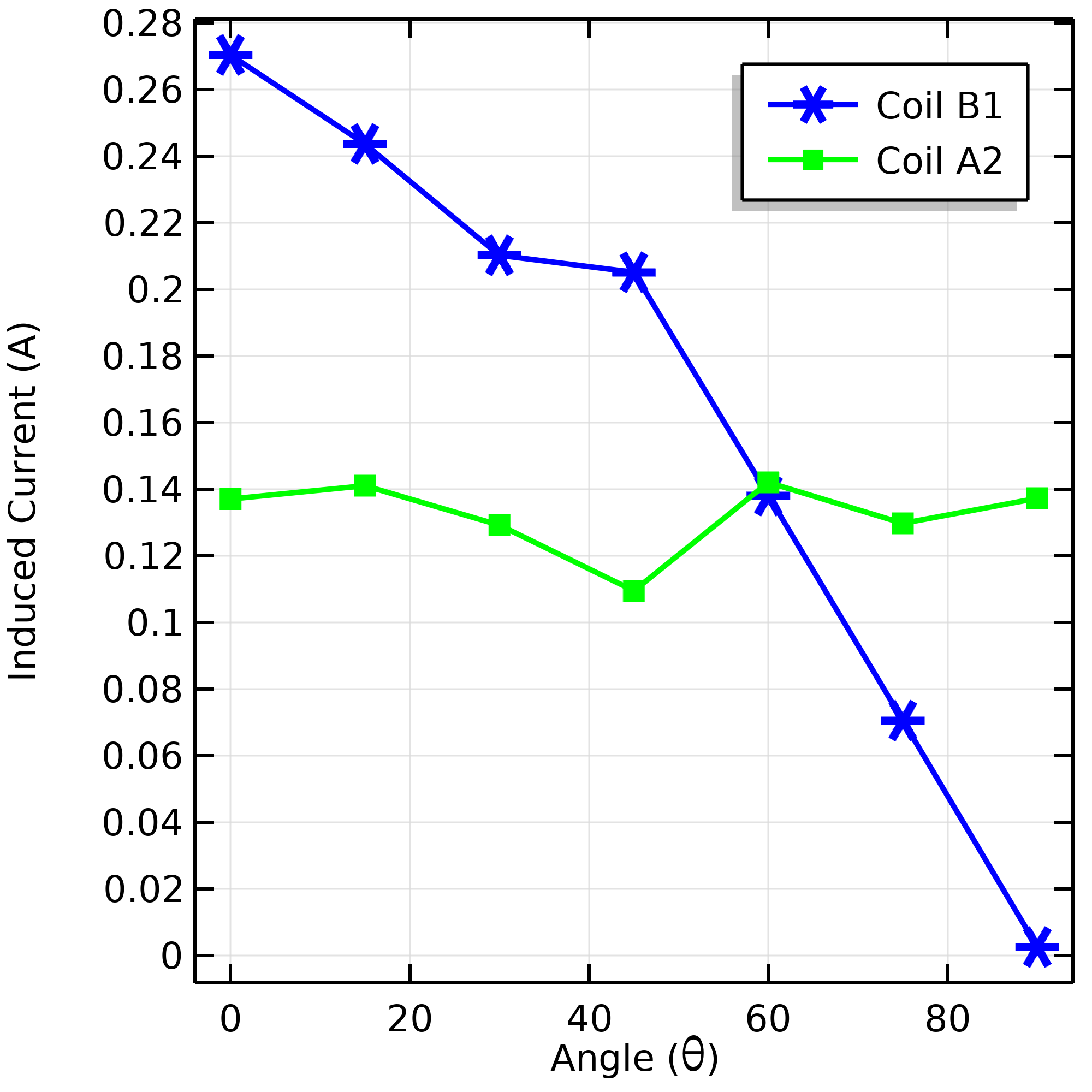} 
    \caption{Orientation sensitivity when Coil A1 and Coil A2 are fixed at $0$cm, and Coil B1 and Coil B2 are placed $67.5$cm away from source coils. The angle of coils B1 and B2 is changed $15^{\circ}$ in a rotational fashion each time with respect to coils A1 and A2.}
    \label{fig:OS_Iind_case3}
    \end{figure} 

\subsubsection{Orientation Sensitivity}
\label{subsec:OS}
Since magnetic fields are directional in nature, it is necessary to evaluate the robustness of any proposed MI architecture/system against changes in transceiver alignment or orientation.
In an underwater environment,  orientation of a sensor node is even more important as underwater sensor nodes are subject to tidal movements. Therefore, in this section, we evaluate our proposed MI FD architecture against rotational misalignment. 

To evaluate, we use the simulation setup such that Node A is kept fixed, while node B is rotated along its axis from $0^{\circ}$ to $90^{\circ}$.
%In FEM simulation environment, the node A coils (Coil A1 and Coil A2) are kept fixed for simplicity, while we keep changing the angle ($0^{\circ}$ to $90^{\circ}$) in a rotational fashion of node B coils (Coil B1 and Coil B2). 
The distance between node A and node B is kept fixed at $67.5$cm in all configurations.
Fig.~\ref{fig:OS_Iind_case3} show interesting results, where the induced current decreases each time when the angle between Coil A1 (Tx) and Coil B1 (Rx) increases from $0^{\circ}$ to $90^{\circ}$. This decreasing trend of induced current in the case of misalignment between Coil A1 and Coil B1 clearly affects the robustness of FD MI communication. However, in case of Coil B2 (Tx) and Coil A2 (Rx), a non-zero induced current at all angles can be observed from Fig.~\ref{fig:OS_Iind_case3}. This is because Coil B2 and Coil A2 are azimuth aligned with each other. However, if the elevation angle between Coil B2 and Coil A2 is changed, then a considerable change can be noted for Coil A2 too.% the induced current needs to be changed accordingly as of Coil B1 in Fig. \ref{fig:OS_Iind_case3}.

%%%%%%%%% SECTION III : Challenges %%%%%%%%%%%
\section{Full-duplex MI Communication: Challenges $\&$ Future Directions}
\label{sec:FD_challenges}
\subsection{Re-configurable Intelligent Surfaces}
\label{subsec:range_ext}
Magnetic field strength decays rapidly (of the order of $1/3$) as nodes move away from the point of origin, hence, limiting the overall communication distance. In the case of the proposed FD MI architecture, the range may further be reduced due to the horizontal TX/Rx (Coil B2/Coil A2) pair as discussed in Section~\ref{subsec:range}. To achieve maximum range in this architecture it is recommended to either increase the transmit current in coil B2 or improve the receive sensitivity of Coil A2. However, increasing the transmit current leads to more power consumption whereas improving the receive sensitivity increases the cost.

Another method to extend the communication distance for MI communication systems is the use of relaying nodes or~\emph{waveguides}. Waveguides are low-power, cost-efficient, and very commonly used in half-duplex MI communication systems~\cite{khalil2020optimal}. However, it will be interesting to study the role and working of waveguides in full-duplex MI communication systems. In particular, the efficiency of the relaying nodes/waveguides to extend the communication two-way needs to be investigated. Furthermore, it will also be interesting to study the effect of self-interference and orientational robustness on the FD MI communication systems with waveguides. 

Alternatively, re-configurable meta-materials based coils can also be used to extend the communication distance~\cite{guo2018full}. Metal-materials or intelligent reflected surfaces (IRS) have the capacity to strengthen the magnetic field in a specific direction and thus can be explored in conjunction with FD MI communication systems.

\subsection{Cross-boundary Communication}
\label{subsec:cross-boundary}
To enable and exploit functionalities of the internet of X-things (IoXT), the communication link between terrestrial and underwater sensor nodes is pivotal~\cite{saeed2019towards}. Since MI communication systems offer similar performance in both air and underwater mediums (due to similar magnetic permeability), a cross-boundary communication link can easily be established. 
Therefore, to equip unmanned air vehicles (UAVs), AUVs or UUVs with FD MI capabilities, two-way cross-boundary transmission is possible without any special gateway nodes. Hence, it is highly recommended to exploit this advantage of MI communication systems to ensure efficient utilization of the network resources in terms of both power and hardware.

%Apart from enabling cross-boundary communication through UAVs/AUVs, UAVs/AUVs can help to overcome the challenge of the limited distance of FD MI system since it can easily move from one point to another.

\subsection{SWIPT Implementation}
\label{subsec:data-power}
Sensor nodes in any wireless sensor network (in general) and underwater wireless sensor networks (in specific) are battery-operated and conventional methods of replacing or charging the batteries are both inconvenient and expensive. Alternate methods are thus required to transfer power to these sensor nodes. One such method is simultaneous wireless information and power transfer (SWIPT). In SWIPT, a sensor node is capable of transmitting both power and information simultaneously~\cite{kisseleff2016magnetic, guo2021joint}. %has been investigated by using different architectures \cite{kisseleff2016magnetic, guo2021joint}. 
SWIPT is very useful in wireless sensor networks and can even be more efficiently used in our proposed FD MI architecture. As each FD MI sensor node contains 2D orthogonal coils, Node A can transfer information and power through Coil A1 and A2, respectively, and the transmitted information and power will be received at Node B through Coil B1 and B2, respectively. 
%The cost of enabling the SWIPT technique in the proposed architecture is a half-duplex link instead of a full-duplex. However, enabling SWIPT through MI can be proven a simple and low-cost alternative as compared to acoustic and optical, especially in the harsh underwater environment.
However, the major challenge with SWIPT implementation in our proposed architecture is its orientation sensitivity, as the misalignment of transceiver coils can affect the efficiency of the system. In the case of an underwater environment, since the alignment of transceiver coils is even more challenging due to the tidal movements,  further investigation and evaluation are needed before deploying the SWIPT technique. 

\subsection{Scalability}
\label{subsec:scalability}
Scalability is the measure of how conveniently sensor nodes can be added to a wireless sensor network~\cite{morozs2020scalable}. %handling of additional sensor nodes in any wireless sensor network and that 
A scalable network does not require major changes in the routing layer, MAC layer, and localization algorithms. %On the other hand, if there needs significant changes in the routing layer, MAC layer or localization algorithms for every single node that enters the network, that network is not scalable and extremely difficult to sustain. 
%Advancements in UWSNs and the internet of underwater things (IoUT) make large-scale networks that will connect tens and hundreds of sensor nodes. Therefore, scalable protocols and localization algorithms are required to adapt easily to the changes in the network.
In general, wireless sensor networks with half-duplex sensor nodes are easily scalable. However, this is not necessarily the case for full-duplex sensor nodes. Similarly, MI sensor nodes are easily scalable with half duplex nodes but not necessarily in the case of full duplex nodes. The proposed architecture, however, provides the opportunity to be scalable in a specific dimension (in which the transceiver coils are aligned to each other). We believe further research in this direction will be influential and rewarding.

\subsection{Security}
\label{subsec:security}
Sensor nodes in Magnetic Induction Wireless Sensor Networks (MIWSN) are generally secure due to the short range, inaudible and invisible nature of magnetic fields~\cite{muzzammil2020fundamentals}. Moreover, the directional nature of the magnetic fields makes the MI wireless sensor network even more secure, as an intruder node can hack the information only when they are perfectly aligned with the legitimate node.

Security of near-field communication is discussed in brief in literature~\cite{rahman2017classification}, however, much more investigation is required to explore how common attacks in a given wireless sensor network may apply to the MI technology. For example, physical layer security (PLS) approaches are often adopted to secure the communication channel between legitimate nodes against adversarial nodes in underwater acoustic communication. Such approaches are not investigated sufficiently in MI technology. Since the underwater MI communication channel is drastically different from the underwater acoustic channel, PLS would involve exploring completely different techniques to achieve authentication, confidentiality, integrity and availability in MI technology.

%%%%%%%% SECTION IV: Conclusion %%%%%%%%
\section{Conclusion and Outlook}
\label{sec:conclusion}
In this paper, we motivate the use of Magnetic Induction in underwater communication and we propose a two-dimensional transceiver architecture that is capable of full-duplex communication. The proposed architecture can efficiently utilize the available spectrum and double the data rate by allowing two-way transmission. Furthermore, we evaluate the proposed architecture's performance against its ability to avoid self-interference, effect on communication distance, and robustness to changes in orientation. Finally, we highlight important research challenges and future directions. The advantages of FD MI communication over FD acoustic communication are low-cost, SI-resistance, ease of synchronization due to low link delays, and seamless cross-boundary two-way communication between air and water. However, the proposed architecture also has a trade-off between data rate and communication distance.

\bibliographystyle{IEEEtran}
\bibliography{ref}
\renewenvironment{IEEEbiography}[1]
  {\IEEEbiographynophoto{#1}}
  {\endIEEEbiographynophoto}
  \vspace{-0.75 cm}
 \begin{IEEEbiographynophoto}
 {Muhammad Muzzammil} received the DoE degree in Information and Communication Engineering from Harbin Engineering University, China in 2021. He received the \textit{Best Poster Award} at The 13th ACM International Conference on Underwater Networks $\&$ Systems (WUWNet’18) and is also being selected in the OES Student Poster Competition. He is currently working as a visiting scholar at Hamad Bin Khalifa University, Doha, Qatar. His research interests lie in the areas of wireless communications, magneto-inductive communication and underwater acoustic communication and networking.
 \end{IEEEbiographynophoto}  
\vspace{-0.75 cm}
 \begin{IEEEbiographynophoto}
 {Saif Al-Kuwari} (Senior Member, IEEE) Dr. Saif Al-Kuwari received a Bachelor of Engineering in Computers and Networks from the University of Essex (UK) in 2006 and two PhD’s from the University of Bath and Royal Holloway, University of London (UK) in Computer Science, both in 2012. He is currently an assistant professor at the College of Science and Engineering at Hamad Bin Khalifa University. His research interests include applied cryptography, Quantum Computing, Computational Forensics, and their connections with Machine Learning. He is IET and BCS fellow, and IEEE and ACM senior member.
 \end{IEEEbiographynophoto} 
 \vspace{-0.75 cm}
 \begin{IEEEbiographynophoto}
 {Niaz Ahmed} (Member, IEEE)  Dr. Niaz Ahmed has received his Ph.D. degree  from Missouri University of Science and Technology, Rolla, USA, in 2017 where he worked in the field of underwater wireless communication and developed low power wireless sensor nodes. Dr. Niaz is one of the pioneers in the field of Magneto-Inductive communication.  He is currently working as an Associate Professor in FAST university. His research interests include embedded systems, wireless communication, underwater wireless sensor network, and magneto-inductive communication systems.
 \end{IEEEbiographynophoto}
 \vspace{-0.75 cm}
  \begin{IEEEbiographynophoto}
 {Marwa Qaraqe} (Senior Member, IEEE) Dr. Marwa Qaraqe is an Associate Professor in the Division of Information and Communication Technology in the College of Science and Engineering at Hamad Bin Khalifa University. She received her bachelor’s degree in Electrical Engineering from Texas A $\&$ M University at Qatar in 2010 and went on to earn her MSc and PhD in Electrical Engineering from Texas A\&M University in College Station, TX, USA in August of 2012 and May of 2016, respectively. Dr. Qaraqe’s research interests cover the following areas: wireless communication, signal processing and machine learning, and their application in multidisciplinary fields, including but not limited to security, IoT, and health. Particular interests are in physical layer security, federated learning over wireless networks, and machine learning for wireless communication, security, and health. Dr. Qaraqe serves a reviewer to several IEEE and technical international journals and conferences, including IEEE Wireless Communication and Networking Conference (WCNC) and IEEE Wireless Communication Magazine and serves on the IEEE Wireless Technical Committee on Wireless Communications. 
 \end{IEEEbiographynophoto}
% that's all folks
\end{document}